\documentclass{jaa}
\usepackage{natbib}
\usepackage{graphicx}

\usepackage{url}

\usepackage[colorlinks=true,citecolor=blue]{hyperref}
\begin{document}\sloppy

\title{A physically-motivated perspective of the Fanaroff-Riley classification of radio galaxies}


\author{Gopal-Krishna\textsuperscript{1}, Paul J.\ Wiita\textsuperscript{2}, Ravi Joshi\textsuperscript{3,*} and Dusmanta\ Patra\textsuperscript{4}}
\affilOne{\textsuperscript{1}UM-DAE Centre for Excellence in Basic Sciences (CEBS), Vidyanagari, Mumbai-400098, India\\}
\affilTwo{\textsuperscript{2}Department of Physics, The College of New Jersey, 2000 Pennington Rd., Ewing, NJ 08628-0718, USA.\\}
\affilThree{\textsuperscript{3}Indian Institute of Astrophysics, Sarjapur Rd., Koramangala, Bangalore-560034, India.\\}
\affilFour{\textsuperscript{4}S.~N.~Bose National Centre for Basic Sciences, Kolkata-700106, India.}

\twocolumn[{

\maketitle

\corres{rvjoshirv@gmail.com}

\msinfo{x January xxxx}{x January xxxx}

\begin{abstract}

A small subset of extragalactic double radio sources, termed HYMORS (HYbrid MOrpholgy Radio Sources), is distinguished by a very unusual, hybrid morphology in terms of  the Fanaroff-Riley (FR) classification. In HYMORS, one radio lobe appears edge-darkened (FR I), while the other shows a well-defined emission peak near its outer edge (edge-brightened, FR II). Such sources are rare but critical for constraining the mechanism responsible for the FR dichotomy, a widely debated issue in extragalactic astrophysics. Here we highlight the need for caution in assigning FR type, in view of some upcoming observational campaigns to confirm HYMORS among the candidates. To illustrate this we highlight the  cases of 3 radio sources which have been perceived to be HYMORS, including the radio galaxy 0500+630 (4C +63.07) which has been claimed to be a good, original example of a HYMORS, with a FR I western lobe and a FR II eastern lobe marked by a prominent terminal hot spot. However, its recent VLASS map at 3 GHz has revealed that the western lobe actually extends much farther out than reported and terminates in a well-defined emission peak. This implies that the source is a regular FR II radio galaxy and not a HYMORS. We also provide a brief perspective of the HYMORS phenomenon and underscore the need to confirm a FR I classification by ruling out additional FR II characteristics, such as an inward lobe-widening and spectral steepening, as well as a lack of prominent radio jet within the lobe.
\end{abstract}


}]


\doinum{12.3456/s78910-011-012-3}
\artcitid{\#\#\#\#}
\volnum{000}
\year{0000}
\pgrange{1--}
\setcounter{page}{1}
\lp{1}

\section{Introduction}
\label{sec:int}
Nearly half-a-century ago, using the aperture-synthesis maps of double radio sources in the 3C catalogue, made with the Cambridge one-mile telescope, Fanaroff \& Riley made a seminal discovery according to which the location of brightness peak within a radio lobe is correlated fairly sharply with the radio luminosity of the source \citep{Fanaroff1974MNRAS.167P..31F}.  Thus, for sources with luminosity $P$ below a critical value\footnote{Using H$_0$ = 70 km s$^{-1}$ Mpc$^{-1}$; the original reference used a different Hubble constant.}, $P_c$ (178 MHz) $\sim 10^{33}$ erg Hz$^{-1}$, the brightness of each lobe was typically found to peak within the first half lobe-length, measured from the parent galaxy, and such lower luminosity sources were termed `Fanaroff-Riley (FR) class I'.  In contrast, for most sources having $P > P_c$, designated as FR II type, the brightness of each lobe was observed to peak in its outer half, usually near the lobe's extremity. The two types of lobes are also referred to as `edge-darkened' (FR I) and `edge-brightened' (FR II). Subsequently, it was surmised that this canonical luminosity devision is actually not so sharp \citep[e.g.][]{Baum1995ApJ...451...88B}. In a further investigation of this basic morphological classification, \citet{Owen1991MNRAS.249..164O} and \citet{Ledlow1996AJ....112....9L} found that $P_c$ itself depends on the optical luminosity of the parent galaxy of the radio source, according to $P_c \propto (L_{opt})^{1.8}$ \citep[for a critique of this proposal, see e.g.,][]{Singal2014MNRAS.442.1656S,Capetti2017A&A...601A..81C,Mingo2019MNRAS.488.2701M}. Moreover, the discovery of low-luminosity FR II sources in the sensitive LOFAR Two-metre Sky Survey (LoTSS) 
has shown that radio luminosity is, by itself, not a reliable predictor of the Fanaroff-Riley morphological class since roughly a tenth of the FR I population lies above the canonical luminosity division at $P_c$ and, conversely, an even higher fraction of the FR II population lies below the luminosity division, which was probably missed out in the original study by \citet{Fanaroff1974MNRAS.167P..31F} simply due to rarity of FR II sources in the local universe \citep{Mingo2019MNRAS.488.2701M, Miraghaei2017MNRAS.466.4346M,Heywood2007MNRAS.381.1093H}.

As is evident from the vast body of literature on this topic, the FR dichotomy continues to be a widely recognised effect, however the underlying cause remains unclear and the remarkable FR correlation has been a subject of much debate \citep[e.g.,][]{Scheuer1996ASPC..100..333S, Tchekhovskoy2016MNRAS.461L..46T, Blackman2022NewAR..9501661B}. As recently  reviewed in \citet{Blandford2019ARA&A..57..467B}, an increasingly supported view is that the weak (initially relativistic) jets that are believed to power FR I lobes, decelerate on kiloparsec scale, becoming trans-sonic and losing their collimation before roughly mid-way between the parent galaxy and the tip of the lobe, and thenceforth their synchrotron plasma propagates out subsonically, like a turbulent plume incapable of terminating in a bright radio spot (e.g., \citealt{Bicknell1984ApJ...286...68B}; see below; also, \citealt{Porth2015MNRAS.452.1089P}). In contrast, the (powerful) jets feeding FR II sources maintain their stability and collimation all the way to near the tip of their respective lobes and are still highly supersonic as they terminate in a compact hot spot, the primary sites of particle acceleration \citep[e.g.,][]{Blanford1974MNRAS.169..395B, Georganopoulos2003ApJ...589L...5G, Kaiser2007MNRAS.381.1548K,Mingo2019MNRAS.488.2701M,Harwood2020MNRAS.491..803H}. At that point, the supersonic bulk relativistic flow of the jet's synchrotron plasma terminates abruptly and the flow gets thermalised at a Mach disk (`working surface'), creating a distinct, compact brightness peak, termed `hot spot' \citep{Hargrave1974MNRAS.166..305H,Miley1971ApL.....8...11M}.   As the Mach disk continues its propagation, the thermalised synchrotron plasma is left behind, or partly flows back towards the parent galaxy and such backflowing (light) synchrotron plasma was predicted to inflate a `spindle-shaped' cavity (`waste energy basket') within the ambient thermal gas, which is observed as a cocoon-like radio lobe surrounding the jet  \citep{Longair1973MNRAS.164..243L,Scheuer1974MNRAS.166..513S,Blanford1974MNRAS.169..395B,Begelman1989ApJ...345L..21B}. It may, however, be noted that the subsonic jets of FR I sources, although incapable of terminating in a hot spot, still appear to be able to sometimes engender a backflow, inflating a broad radio lobe \citep[][and references therein]{Laing2012MNRAS.424.1149L, Ruiter1990A&A...227..351D}, as also indicated by some numerical simulations of light relativistic jets  \citep[e.g.,][]{Krause2005A&A...431...45K,Perucho2007MNRAS.382..526P,Rossi2008A&A...488..795R}.

In some FR II sources, the terminal hot spot might appear as no more than a `warm spot' (i.e., a modest local brightness enhancement), for instance due to (i) its radiation being beamed away from us 
\citep[e.g.,][]{Tribble1992MNRAS.256..281T,Kharb2015IAUS..313..211K}, or (ii) a    combination of modulating power of the central engine and differential light travel time between the approaching and receding parts of the radio source, which can shape the observed radio morphology  \citep[e.g.,][]{Gopal1996A&A...316L..13G,Marecki2012A&A...544L...2M,Kapinska2017AJ....154..253K,Harwood2020MNRAS.491..803H}. In such situations, one might look for alternative markers of a Mach disk and the associated backflow of synchrotron plasma \citep[e.g .,][]{Cotton2020MNRAS.495.1271C}.   These proxies include (i) an inward steepening of radio spectrum along the lobe axis \citep{Leahy1989MNRAS.239..401L,Witta1990ApJ...353..476W,Carilli1991ApJ...383..554C,Gawronski2006A&A...447...63G,Kapinska2017AJ....154..253K,de-Gasperin2017MNRAS.467.2234D}, (ii) a magnetic field vector normal to the putative jet direction near the lobe's tip, revealing compression of the synchrotron plasma thermalised downstream of the advancing Mach disk, due to the impact of the external medium \citep[e.g.,][]{Laing1981MNRAS.195..261L}, and (iii) a lobe widening towards the parent galaxy 
\citep[`spindle shape'; cf.][]{Scheuer1974MNRAS.166..513S,Nath1995MNRAS.274..208N}.  Such markers provide supportive signatures of the terminal hot spot, a key feature of FR II lobes, even if the `hot spot' itself appears as no more than a warm spot.

\section{Approaches to the FR Dichotomy}

Possible reasons for the disruption of relativistic jets have been debated for almost four decades and it is fair to state that consensus is tilting in favour of the `extrinsic scenario' in which jet-environment interaction plays a key role in determining the lobe's morphology. Broadly, two main flavours of the interaction are considered. Firstly, the transition of a low-power jet from collimated laminar to a plume-like flow could occur when the jet becomes unstable due to (i) turbulence induced by plasma instabilities, including those resulting from surface entrainment of ambient thermal gas \citep[e.g.,][]{Bicknell1984ApJ...286...68B,De-young1993ApJ...405L..13D,Komissarov1994MNRAS.269..394K,Laing2002MNRAS.336.1161L}, or (ii) by thermal mass loading, e.g., due to stellar winds 
\citep[e.g.,][]{Komissarov1994MNRAS.269..394K,Bowman1996MNRAS.279..899B,Hubbard2006MNRAS.371.1717H,Laing2002MNRAS.336.1161L,Kaiser2007MNRAS.381.1548K,Perucho2007MNRAS.382..526P,Wykes2015MNRAS.447.1001W}. Such jet-stellar wind interactions in AGN have also been invoked to explain their $\gamma$-ray emission 
\citep[e.g.,][]{Bednarek1997MNRAS.287L...9B,Araudo2010A&A...522A..97A,Barkov2012ApJ...749..119B}.

The second framework for understanding the decollimation of light jets into a plume-like flow, even though they start out relativistic and supersonic \citep[e.g.,][]{Biretta1995ApJ...447..582B,Venturi1995ApJ...454..735V,Hardcastle2003ApJ...593..169H}, is via a weakening of the hot spot as its advance speed decreases and approaches the sound speed of the ambient medium. The ensuing loss of ram pressure causes weakening of the (outer)  bow shock, accompanied by a decline in the pressure of the thermal ambient medium crossing the bow shock. This leads to destabilization of the 'contact discontinuity' between the bow shock-heated ambient thermal plasma and the synchrotron plasma in the hot spot region (which is confined by the former, according to the standard model for FR II sources, e.g., \citealt{Blanford1974MNRAS.169..395B}.  Such a possibility of fading/vanishing of the hot spot when its motion is no longer supersonic, has been considered by several authors (e.g., \citealt{Gopal-Krishna1987MNRAS.226..531G,Blandford1996cyga.book..264B,Kawakatu2009ApJ...697L.173K}, also \citealt{Konar2013MNRAS.436.1595K}). The idea that the presence of a hot spot is indicative of its supersonic motion has also been a part of the earliest theoretical models of classical double radio sources \citep[see, e.g.,][]{Blanford1974MNRAS.169..395B}

Radically different descriptions of potentially greater fundamental import have also been put forth to explain the FR dichotomy. According to one such proposal, radio sources belonging to the two FR classes may be powered by fundamentally different types of relativistic VLBI jets: pair plasma jets in the FR I type and electron-proton jets in the FR II type sources 
\citep[e.g.,][]{Celotti1993MNRAS.264..228C,Reynolds1996MNRAS.283..873R,Celotti1997MNRAS.286..415C,Meliani2010IJMPD..19..867M,Croston2018MNRAS.476.1614C}.  Likewise, differences in intrinsic properties of the central engine (e.g., spin of the BH, or the accretion mode) have been invoked as explanations for the FR dichotomy \citep[e.g.,][]{Baum1992ApJ...389..208B, Baum1995ApJ...451...88B,Meier1999ApJ...522..753M,Wold2007A&A...470..531W}. In some other descriptions,  different scenarios involving effects of dust and galactic mergers affecting the accretion onto the central engine, have been envisaged 
\citep[e.g.,][]{Muller2004A&A...426L..29M,Saripalli2012AJ....144...85S}

Although \citet{Fanaroff1974MNRAS.167P..31F} classified radio sources on the basis of whether the separation between the points of peak intensity were less than (FR I) or greater than (FR II) half of the total extent of the radio source, more subtle and physically-motivated FR classification criteria have since been invoked  \citep{Gopal2000A&A...363..507G}. These criteria are: (a) whether the extended lobe emission is best described as plume-like or bridge-like; and (b) whether the lobes possess compact features or not, and if they do, whether the compact emission is dominated by (terminal) hot spots, weak jets or strong jets \citep[e.g.,][]{Leahy1993LNP...421....1L}.  In some rare instances, even though a terminal hotspot is not identifiable, its existence can be inferred if a spectral flattening is observed near the lobe extremity \citep[e.g., the source B2 0924+35, see][]{Morganti2021A&A...648A...9M}. Even if a terminal hotspot is observed in a lobe but it is not the lobe's brightest part, so long as the hot spot is well defined, the basic tenet of FR II classification is satisfied, because such a hot spot is an evidence that the jet has remained collimated and supersonic until the lobe's extremity. Such a situation may well arise in the case of `restarted' double radio sources, which no longer seem to be so rare \citep[e.g.,][]{Jurlin2020A&A...638A..34J}. Among these, the most pursuasive examples of episodic jet activity are the so called `double-double radio sources' \citep[e.g.,][]{Schoenmakers2000MNRAS.315..371S,Kaiser2000MNRAS.315..381K,Stawarz2004ApJ...613..119S, Safouris2008MNRAS.385.2117S,Saikia2009BASI...37...63S,Konar2013MNRAS.436.1595K,Brienza2018A&A...618A..45B}. In these highly evolved old sources, the terminal hot spots of the outer lobes may often be viewed in a fading phase, whereas the much younger inner lobes can be bright. Several such examples have been reported in the literature, showing the inner lobes extending on kpc-scale \citep[e.g.,][]{Brocksopp2007MNRAS.382.1019B,Hota2011MNRAS.417L..36H,Sing2016ApJ...826..132S,Bruni2020MNRAS.494..902B} to $10^2$ kpc scale  \citep[e.g.,][]{Saripalli2003ApJ...590..181S,Mahatma2019A&A...622A..13M,Dabhade2020A&A...642A.153D,Bruni2021A&A...654A..27B,Lal2021ApJ...915..126L}. Such brightness peaks within the lobes (due to episodic activity) may thus be mistaken as signifying FR I morphology. However, one expects them to be distinguishable via a localised radio spectral flattening \citep[e.g.,][]{Saikia2009BASI...37...63S}. Thus, in essence, the detection of a well-defined radio peak (hot spot) near the lobe's extremity, even if it is not the brightest part of the lobe, emerges as a reliable criterion for FR II classification.

\section{The discovery of HYMORS}

The afore-mentioned range of prescriptions seeking to relate the FR dichotomy fundamentally to processes other than the jet's interaction with ambient medium (ahead or around the jet, see Sect.\ 2), such as a difference between the composition of their jets, or the spin of their black holes, or in the process of accretion on to the central engine, were challenged and called into question by the discovery of `HYbrid MOrphology Radio Sources' (HYMORS), i.e., double radio sources in which clearly different FR morphologies are observed for the two lobes
\citep{Gopal2000A&A...363..507G,Gopal2001A&A...373..100G,gopal2002NewAR..46..357G}. Although HYMORS are rare objects and only a few dozen cases have so far been documented 
\citep[e.g.,][]{Gopal2000A&A...363..507G,Gawronski2006A&A...447...63G,Kharb2015IAUS..313..211K,de-Gasperin2017MNRAS.467.2234D,Kapinska2017AJ....154..253K,Dabhade2020A&A...635A...5D,Harwood2020MNRAS.491..803H,Kumari2022MNRAS.514.4290K}, a  majority of them still being only candidates in need of confirmation, it is widely recognised that they can play a crucial role in pinning down the cause of the FR dichotomy (Sect.\ 2). Therefore, an increasing amount of effort is being invested in observing campaigns to select bona-fide HYMORS out of the candidates. Bearing this in mind, we deem it prudent to draw attention to some basic factors, neglect of which may lead to a spurious FR classification of radio lobes. Below, we illustrate this by discussing one specific case where the suggested FR I classification of a radio lobe does not seem well-founded, in the light of its recent higher quality radio map. In addition, we shall highlight another two sources which have been clearly mis-classified as likely HYMORS, based on radio maps lacking in angular resolution and/or sensitivity.

\begin{figure*}
\begin{center}
\includegraphics[width=16 cm]{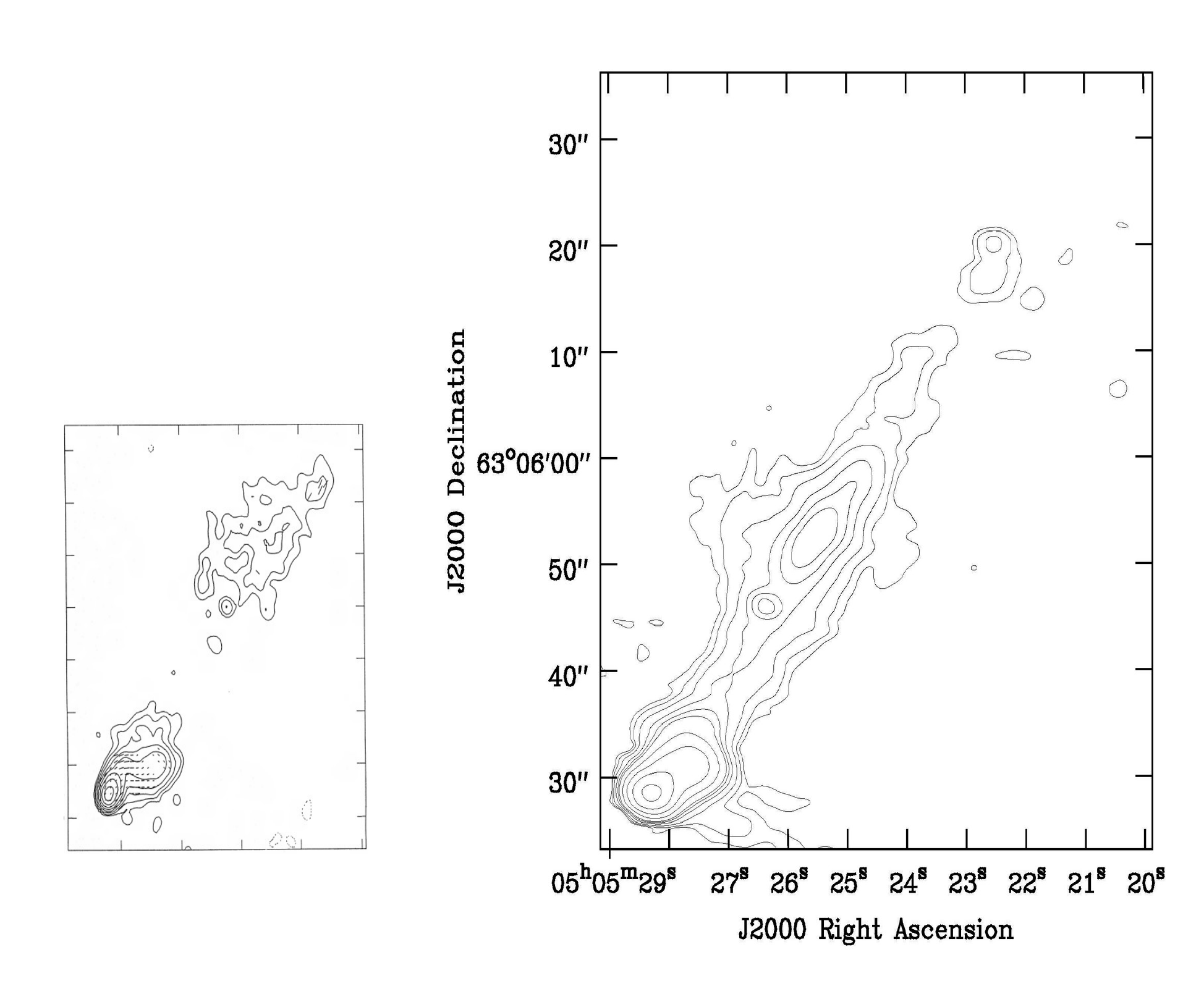}
\caption{(Left) VLA map of 0500+630 at 1.665 GHz, reproduced from \citet{Saikia2022arXiv220605803S}. The contour levels are at $1.00\times 10^{-3}$Jy $\times$ (-1, 1, 2, 4, 8, 16, 32, 64, 128, 256, 512, 1024, 2048, 4096), and the beamwidth is  $1.3^{\prime \prime}\times  1.0^{\prime \prime}$. (Right) VLASS map with a beamwidth of $3.5^{\prime \prime}\times  2.2^{\prime \prime}$ at 3.0 GHz; the contour are drawn at $2.7 \times 10^{-4}$ Jy $\times$ (2, 4, 8, 16, 24, 32, 64, 128, 256, 512). The radio core, seen in both maps, is located at RA = 05 05 26.32 and Dec = +63 05 45.9 (J2000). The two map have identical scales and are aligned in declination. 
}\label{Fig1}
\end{center}
\end{figure*}

\begin{figure*}
\begin{center}
\includegraphics[width=18 cm]{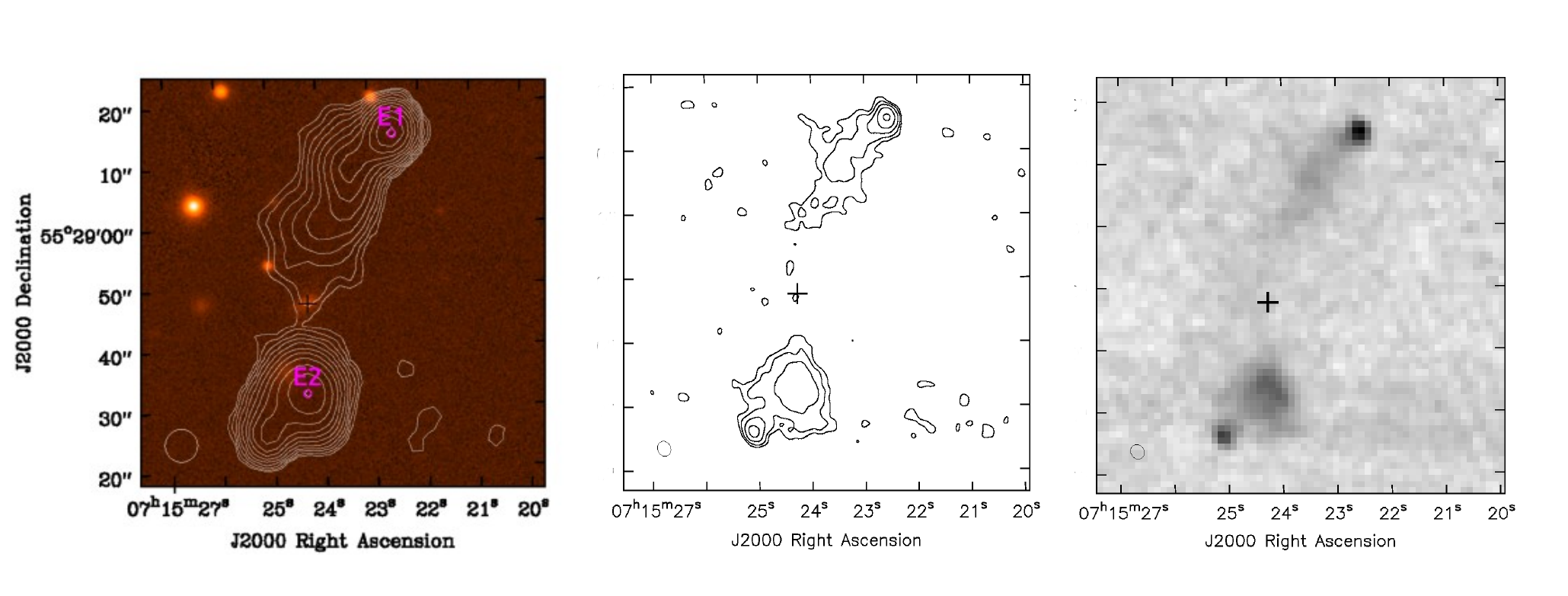}
\includegraphics[width=18 cm]{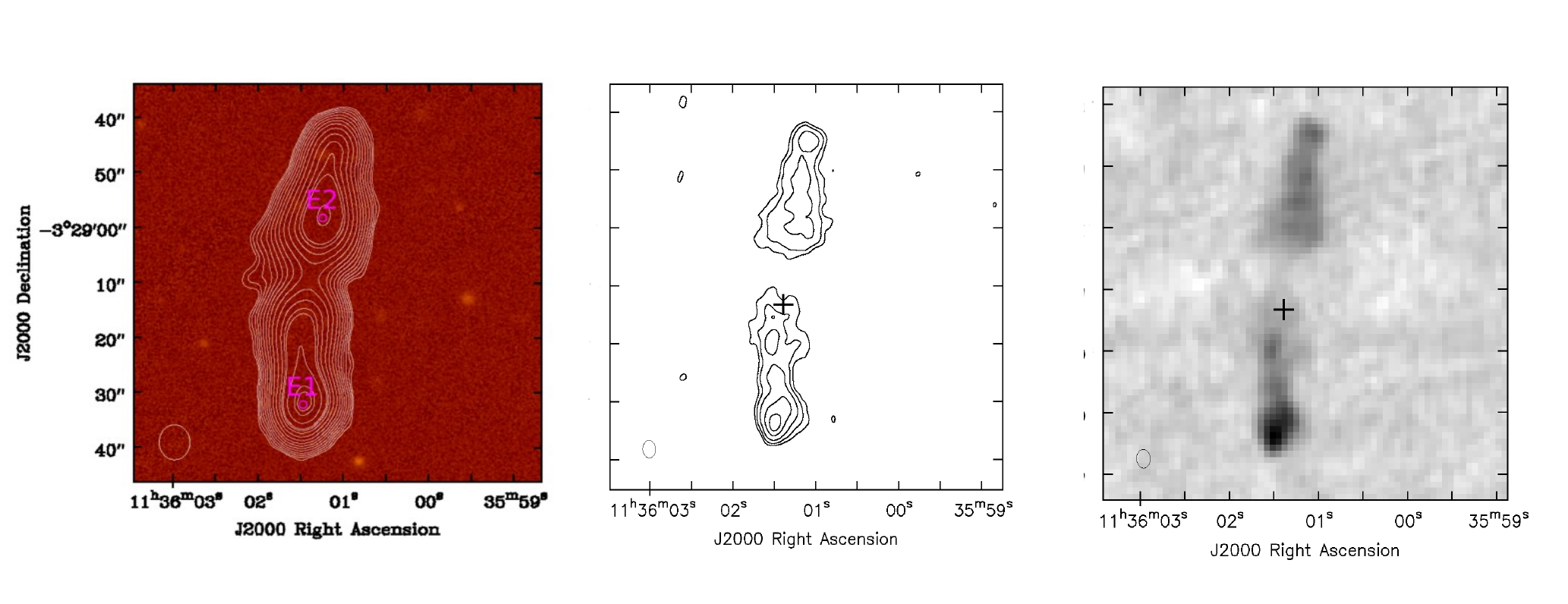}
\caption{ Left Panels: 1.4 GHz FIRST maps (beamsize $5^{\prime \prime}$) of the two HYMORS candidates J0715+5528 and J1136-0329, reproduced from 
 ``Search for hybrid morphology radio galaxies from the FIRST survey at 1400 MHz'', Fig. 2 of \citet{Kumari2022MNRAS.514.4290K}.
The middle and the right panels show the 3 GHz VLASS maps (beamsize $\sim 3.03^{\prime \prime} \times 2.29^{\prime \prime}$) of the same two sources, in contour and grey-scale representations \citep{Lacy2020PASP..132c5001L}, where the plus signs indicate the location of the host galaxy.
}\label{Fig2}
\end{center}
\end{figure*}

\section{Discussion}

The first source is the radio galaxy 0500+630 (4C +63.07, J050526+630546), claimed to be a HYMORS identified with a 18-mag galaxy at $z = 0.290 \pm 0.004$ and having an flux density of $\sim$ 5.5 Jy at 178 MHz, whose VLA map at 1.665 MHz is displayed in Fig.\ 1 \citep{Saikia1996MNRAS.282..837S}, together with its VLASS (3 GHz) map, both maps drawn to identical scales. A compact radio core is seen to coincide with the galaxy. A striking aspect of these maps is the enormous surface-brightness contrast between the two lobes, as originally highlighted by \citet{Saikia1996MNRAS.282..837S}. Since, in their 1.665 GHz VLA map, a well-defined brightness peak was not discernible near the tip of the western lobe, they designated the lobe as FR I type. This, together with the fact that the lobe on the opposite (eastern) side exhibits a prominent terminal hot spot and hence is edge-brightened (FR II), led them to classify this source as a HYMORS. If true, this would be the first HYMORS to be recognized, as indeed mentioned, e.g., in
\citet{Kapinska2017AJ....154..253K,Dabhade2020A&A...635A...5D,Bruni2021MNRAS.503.4681B}, and \citet{Saikia2022arXiv220605803S}. 

On the other hand, the lack of a well-defined terminal hot spot in the western lobe could even be due to the beaming-away of the radiation from the putative hot spot, or due to its fading arising from a decline in the power of the central engine, coupled with differential light travel time for the two lobes (Sect.\ 1). Both these possibilities would be consistent with a FR II designation for the western lobe. In fact, this is corroborated by the observed shape of the western lobe, which is seen to gradually widen going inwards, exhibiting the `spindle-shaped' boundary which is a characteristic of FR II lobes, originally predicted by \citet{Scheuer1974MNRAS.166..513S}. Further validation of this interpretation would come from spectral index mapping along this lobe, recalling that a spectral steepening towards the nucleus is typically expected for a FR II lobe (Sect.\ 1). However, a spectral index map of this source is not available at present.  

An unexpected outcome emerging from a comparison of the two maps shown in Fig.\ 1 is that the outer half of the western lobe is missing in the VLA map reported in  \citet{Saikia1996MNRAS.282..837S} and \citet{Saikia2022arXiv220605803S}! The missed-out outer part, which is captured in the VLASS map, in fact, does show a well-defined brightness peak near the extremity of the western lobe (Fig.\ 1). This peak can be readily identified as the terminal Mach disk of the western jet, testifying that, like its eastern counterpart, the jet feeding the western lobe too has remained collimated and supersonic until the tip of the lobe. A radio trail is seen extending from the terminal hot spot towards the inner part of the western lobe, affirming that the hot spot is physically connected to the rest of the western lobe. This is further bolstered by the observed nearly perfect alignment of the western hot spot with the axis passing through the bright eastern hot spot and the radio nucleus (Fig.\ 1). The two terminal hot spots are separated by 64.7$^{\prime \prime}$, which corresponds to an overall projected size of 291 kpc for this double radio source, in a flat cosmological model with $H_0 = 67.4$ km s$^{-1}$ Mpc$^{-1}$, $\Omega_m = 0.315$, and $\Omega_{\Lambda} = 0.685$ \citep{Planck_Collaboration2020A&A...641A...6P}.
The lengths of the western and eastern lobes are 193 kpc and 98 kpc, implying an unusually large lobe-length ratio of $\sim 2.0$, validating the claim of \citet{Saikia1996MNRAS.282..837S} about this source being highly asymmetric (in terms of surface-brightness of the two lobes). Two other inferences stemming from the VLASS radio map are:
 
(i) The clear detection of a terminal hot spot in the western lobe, together with its inward-widening shape (Sect.\ 2; Fig. 1) are consistent with the current conceptualisation of FR II lobes (Sect.\ 1). Thus, with both its lobes being of FR II type, this radio galaxy is a regular FR II source, not a HYMORS.

(ii) The updated radio structure of this source is now clearly in accord with the `Mackay's rule' for FR II radio sources, according to which the brighter of the two terminal hot spots appears closer to the parent galaxy \citep{Mackay1971MNRAS.154..209M}. A comprehensive discussion on this and a range of other issues related to asymmetries in radio sources has been reviewed in \citet{gopal2004astro.ph..9761G}.

 To further highlight this issue, we now consider the two sources, J0715+5528 and J1136-0329, recently claimed to be HYMORS candidates \citep{Kumari2022MNRAS.514.4290K}, as mentioned in Sect. 3. The claim was made on the basis their radio maps from the NVSS \citep{Condon1998AJ....115.1693C} and FIRST \citep{Becker1995ApJ...450..559B} surveys. Redshift is available only for J1136-0328 ($z = 0.82$) \citep{Kumari2022MNRAS.514.4290K}. In Fig.\ 2 we present overlays of their Pan-STARRS-FIRST maps, reproduced from \citet{Kumari2022MNRAS.514.4290K}, alongside their VLASS images in both contour and gray-scale representations, made with a $3.03^{\prime \prime} \times 2.29^{\prime \prime}$ beam at 3 GHz \citep{Lacy2020PASP..132c5001L}. These VLASS images leave no doubt that in each source, both lobes contain a terminal hot spot, again demonstrating that both are canonical FR II sources and not HYMORS.

\subsection{Tenability of FR classification}

Coming back to the radio galaxy 0500+630, its FR II classification inferred here is also in accord with its radio luminosity and optical spectrum.  The source has a radio luminosity of  $6 \times 10^{33}$ erg s$^{-1}$ Hz$^{-1}$ at 178 MHz which, as also mentioned by \citet{Saikia1996MNRAS.282..837S}, is well above the luminosity ($P_c)$ which is supposed to divide the two FR classes. They have further pointed out that its parent galaxy has an optical spectrum exhibiting a strong [O III] line, as well as a high (H$\alpha$ + [N II]) / H$\beta$ ratio of $\sim 10$. Such a high-excitation spectrum is a rarity for FR I radio galaxies \citep[e.g.,][]{Mingo2019MNRAS.488.2701M} and is indicative of an ongoing activity of the central engine and feeding of both hot spots by the jets, as envisaged here. \par

In conclusion, it can  be surmised from the examples presented here that FR classification of a radio lobe purely on the basis of relative location of the brightness peak within the lobe can be potentially misleading \citep[see, also,][]{de-Gasperin2017MNRAS.467.2234D}. For a secure classification of a suspected FR I lobe, the possibility of its being a FR II lobe must be firmly excluded. This would require a careful check for the presence of at least some of the well-known features symptomatic of FR II lobes. Aside from the presence of a well-defined brightness peak near the lobe's extremity, the other FR II attributes include (Sect.~1): a magnetic field orientation roughly orthogonal to the inferred jet direction near the lobe's extremity, a widening profile of the lobe's boundary and radio spectral index steepening towards the  nucleus, and the presence of a prominently visible jet within the lobe. In this quest, multi-frequency radio maps of high fidelity and with polarimetric information can play a key role. The imprints of orientation effects in simulating HYMORS characteristics \citep{Harwood2020MNRAS.491..803H} would be another important direction of inquiry which will need checks based on multi-band observational inputs.\\ \\

\vspace{-1cm}

\section*{Acknowledgements}
We thank the reviewers for their constructive comments.
GK acknowledges award of a Senior Scientist fellowship of the Indian National Science Academy. DP acknowledges the post-doctoral fellowship of the S.~N.~Bose National Centre for Basic Sciences, Kolkata, India, funded by the Department of Science and Technology (DST), India.
This research has made use of the CIRADA cutout service at URL cutouts.cirada.ca, operated by the Canadian Initiative for Radio Astronomy Data Analysis (CIRADA). CIRADA is funded by a grant from the Canada Foundation for Innovation 2017 Innovation Fund (Project 35999), as well as by the Provinces of Ontario, British Columbia, Alberta, Manitoba and Quebec, in collaboration with the National Research Council of Canada, the US National Radio Astronomy Observatory and Australia's Commonwealth Scientific and Industrial Research Organisation.
\vspace{-1em}

\bibliography{ms_HYMORS_JAA_rev}

\end{document}